\documentclass[doublecol]{epl2} 
\usepackage{graphicx}
\usepackage{epsfig}
\usepackage{amssymb,amsmath}
\usepackage{bm}

\def\gsim{\:\raisebox{-0.5ex}{$\stackrel{\textstyle>}{\sim}$}\:}
\def\3dots{\:\raisebox{-0.5ex}{$\stackrel{\textstyle.}{:}$}\:}
\def\beq{\begin{equation}}
\def\eeq{\end{equation}}
\def\bea{\begin{eqnarray}}
\def\eea{\end{eqnarray}}

\title{Buckling and force propagation along intracellular microtubules}
\shorttitle{Buckling and force propagation along intracellular microtubules}

\author{Moumita Das\inst{1,2} \and Alex J. Levine\inst{3} \and F.C. MacKintosh\inst{1,2}}
\shortauthor{M. Das \etal}

\institute{                    
  \inst{1} Department of Physics and Astronomy, Vrije Universiteit, Amsterdam, The Netherlands.\\
    \inst{2} The Aspen Center for Physics, Aspen, CO 81611.\\
    \inst{3}  Department of Chemistry and Biochemistry, and University of California, Los Angeles, CA 90095.\\
    California Nanosystems Institute, University of California, Los Angeles, CA 90095.
    }

    \abstract{ Motivated by recent experiments showing the compressive buckling of microtubules in cells, we study theoretically the mechanical response of, and force propagation along elastic filaments embedded in a non-linear elastic medium. We find that embedded microtubules buckle when their compressive load exceeds a critical value $f_c$,  and that  the resulting deformation is restricted to a penetration depth that depends on both the non-linear material properties of the surrounding cytoskeleton, as well as the direct coupling of the microtubule to the cytoskeleton. The deformation amplitude depends on the applied load $f > f_c$ as $(f- f_c)^{1/2}$. This work shows how the range of compressive force transmission by microtubules can be tens of microns and is governed by the mechanical coupling to the surrounding cytoskeleton. }

\begin{document}

\maketitle

The mechanical response of most eukaryotic cells depends on their \emph{cytoskeleton}, a composite network of filamentous proteins ~\cite{alberts}. 
Microtubules (MTs) are the stiffest of these cytoskeletal filaments, and they play an important role in organization of, and transport within the 
cell. Their mechanical rigidity allows them to support significant stresses in the cytoplasm. These stresses can be highly inhomogeneous, 
with compressive/tensile forces directed along stiff MTs, permitting directed force transmission and mechanical signaling over several 
microns within the cell. As with macroscopic elastic rods, however, even the comparatively rigid MTs cannot, on their own, withstand as large 
\emph{compressive} loads as \emph{tensile} loads. This is because of the classical Euler buckling instability limiting the compressive force to a 
maximum value, which actually vanishes for long rods. It was recently shown, however, that even long MTs \emph{can} bear large compressive 
loads, as a result of their coupling to the surrounding elastic matrix of the cytoskeleton \cite{cliff}. This composite aspect of the 
cytoskeleton has important consequences for cell mechanics and \emph{mechanotransduction} \cite{ingber, janmey:2004, janmey:1998, 
wang}---the generation, transmission, and sensing of forces by the cell.

Here, we develop a model for compressively loaded elastic filaments such as MTs embedded in an elastic continuum. When their compressive load $f$ exceeds a critical force $f_c$, an oscillatory buckling of the filament is expected, with a wavelength depending on both the stiffness of the elastic filament and the shear modulus of the surrounding medium~\cite{cliff,landau,skotheim}. In the classical Euler buckling problem, and even in the presence of a (linear) elastic background, an elastic rod becomes unstable for $f>f_c$. We include the non-linear elastic properties expected for the cytoskeleton, and show that the system is stable to supercritical loads, with a buckling amplitude that increases above threshold as $|f-f_c|^{1/2}$. In addition, both the buckling amplitude and the compressive load decay away from the point of force application in a way that depends sensitively on the longitudinal mechanical coupling of the filament to its surroundings. This suggests that the range of force transmission in the cell can be effectively controlled by microtubule-associating proteins that couple MTs to the rest of the cytoskeleton. Surprisingly, the experiments in Ref.\cite{cliff} (see Fig.6 therein) also found evidence that, despite the existence of a threshold force for buckling, the deformation of the microtubule was attenuated, suggesting a spatially varying force.

In studying the mechanical response of intracellular MTs, we extend the classical buckling theory of rods in an elastic medium \cite{cliff,landau,skotheim} to take into account the mechanical coupling of both longitudinal and transverse deformations of the MT to the surrounding \emph{nonlinear} elastic cytoskeletal network. We use a variational theory, numerical (conjugate gradient) minimization, and asymptotic analysis of the deformation of the MT. On applying a compressive load at one end, the microtubule bends and deforms the surrounding elastic network which resists this deformation. These competing effects determine the characteristic force and length scales associated with the buckling of the MT. The nonlinear elastic properties and longitudinal coupling of the MT to its surroundings determine the attenuation of both the force and deformation of the MT along its arclength away from the point of loading. While the details of the deformation profile depend on the precise form of the nonlinearity, the force penetration depth generically decreases with increasing strength of the longitudinal coupling between the MT and the medium.

We consider a linear elastic theory for the MT and the surrounding medium and include the leading nonlinearity for the elasticity of the surrounding matrix. The MT is subjected to a compressive load $f$ at one end denoted by $x=0$.  The resulting elastic deformation energy can be written in terms of its transverse and longitudinal displacements, $u(x)$ and $v(x)$, respectively:
\bea
\label{freeenergy}
\mathcal{E} = - f v(0) + \int _0^{\infty} \left [ {  \frac{\kappa}{2} {u^{''}(x)}^2} +  \frac{\alpha_{\perp}}{2} {u(x)}^2 \right ]  dx  \nonumber \\
+  \int _0^{\infty} \left [ { \frac{\alpha_{\parallel}}{2} {v(x)}^2  +  \frac{ \beta}{4} {u(x)}^4} \right ] dx,
\eea
where  $v(x) = \int_x^{\infty} \!\! \frac{1}{2}{u^{'}(y)}^2 dy$ as is required for
an {\em incompressible} rod. The first term represents the axial compression energy released by bending. The second term corresponds to the bending energy of the rod, $\kappa$ is the bending rigidity. Classical Euler buckling, in which a long-wavelength bend of a rod of length $L$ occurs at a threshold force of $\pi^2\kappa/L^2$, results from just the first two terms \cite{landau, dogterom}. The elastic constants  $\alpha_{\perp}$, $\alpha_{\parallel}$ correspond to the linear elastic transverse and longitudinal mechanical couplings between the MT and the surrounding actin network. The transverse coupling is directly related to the (linear) shear modulus $G$ of the surrounding medium: $\alpha_\perp=4\pi G/\ln(\lambda/a)$, where $\lambda$ is the characteristic wavelength of lateral displacement and $a$ is a small wavelength cut-off. Adding just this third term leads to buckling on a short wavelength $\lambda= 2 \pi \ell_0$, where $\ell_0=(\kappa/\alpha_{\perp})^{1/4}$ \cite{landau,cliff}, and at a force threshold $f_c= 8\pi^2\kappa/\lambda^2=2  (\kappa \alpha_\perp)^{1/2}$ that is significantly larger than for Euler buckling. The longitudinal coupling, in contrast, is not directly related to the elasticity of the surrounding medium, but also depends on the strength/degree of attachment between the MT and the surrounding network. Thus, it falls in the range $0\leq\alpha_\parallel\leq2\pi G/\ln(\lambda/a)$. Here, the lower limit corresponds to an uncoupled MT that slides freely along its axis, while the upper limit corresponds to the strong coupling of the MT to the medium along its length. Finally, the term proportional to $\beta > 0$ represents the leading non-linear transverse coupling to the medium. In fact, both terms in the second integrand of Eq.\ (\ref{freeenergy}) are non-linear in the transverse displacement $u$ since the first of these is effectively fourth-order in $u$, due to the quadratic relationship between $u$ and $v$.

\begin{figure}
\onefigure[width=8cm]{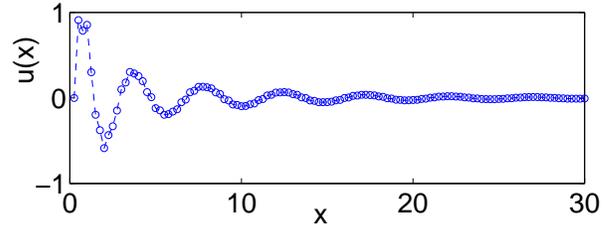}
\caption{The deformation of a stiff elastic filament embedded in a nonlinearly elastic medium decays as an oscillatory exponential (from numerics). The material parameters are  $\kappa=\beta=\alpha_{\perp}=1.0$, and $\alpha_{\parallel}=0.1$.}
\label{Fig.1}
\end{figure}

\begin{figure}
\onefigure[width=8cm]{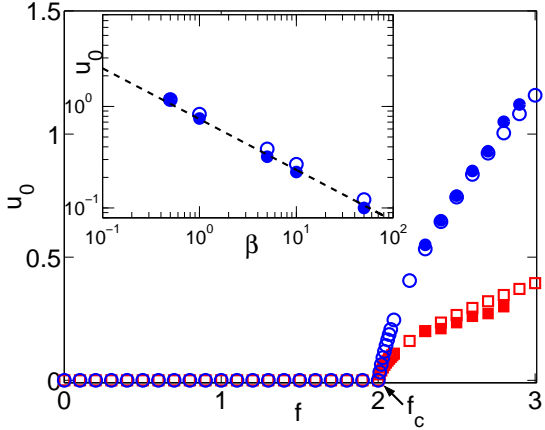}
\caption{The amplitude $u_0$ as  function of  compressive load $f$ for
$\kappa=\alpha_{\perp}=1.0, \alpha_{\parallel}=0.1$, at $\beta=1$ (circles) and $\beta=10.0$ (squares). Inset shows the amplitude as a function of $\beta$, with  $f/f_c=1.25$. In both the main figure and the inset, filled symbols correspond to the numerics, open symbols to the variational calculation. The dashed line represents the prediction of the asymptotic analysis.}
\label{Fig.2}
\end{figure}

\begin{figure}
\onefigure[width=8cm]{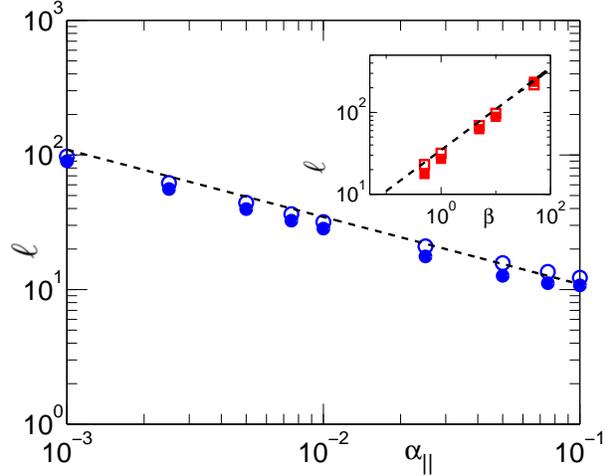}
\caption{Decay length $\ell$ as a function of the longitudinal coupling constant $\alpha_{\parallel}$ for $\kappa=\beta=\alpha_{\perp}=1.0, f/f_c=1.25$.  Inset shows the variation of the decay length with the nonlinear elastic constant $\beta$ for $\kappa=\alpha_{\perp}=1.0$ and $\alpha_{\parallel}=0.01$ and $f/f_c=1.25$. Filled symbols in both the main figure and inset  correspond to the numerics, open symbols to the variational calculation and the dashed lines to the asymptotic analysis.}
\label{Fig.3}
\end{figure}

We examine the effect of the nonlinear terms in Eq.~(\ref{freeenergy}) by numerically determining the deformation of a MT in the surrounding matrix that is subjected to symmetric compressive loads applied at both ends: $x=0$ and $x=L$. In the limit of large $L$, the solution near $x=0$ is equivalent to that of a MT loaded at just one end, and the symmetrized loading is simply for computational convenience. Using a conjugate gradient technique~\cite{press}, we minimize the energy of the loaded system given by Eq.~(\ref{freeenergy}) with an additional term $+f v(L)$ representing the load at $x=L$ and subject to hinged boundary conditions. We find that the buckling threshold $f_c$ and wavelength $\lambda = 2 \pi \ell_0$
remain approximately the same as in the linear system, but that the amplitude of the undulatory deformation $u(x)$ just above
threshold (i.e., for small $f - f_c>0$) is highly attenuated (see Fig.\ \ref{Fig.1}). The buckling amplitude scales with the applied force $f$ as  $(f - f_c)^{1/2}$ reminiscent of a supercritical pitchfork bifurcation~\cite{Crawford:91}; it also scales with the nonlinear elastic constant $\beta$ as a power law with exponent $-1/2$ as shown in the inset of Fig.\ \ref{Fig.2}. At a given force, the oscillatory deformation field decays with an apparently exponential envelope as one moves along the axis of the MT away from the loading point, with a penetration depth or decay length that scales as $({\beta/\alpha_\parallel})^{1/2}$ (Fig.\ \ref{Fig.3}).

We propose a variational ansatz for the form of the transverse deformation field based on
the numerical solutions shown in Fig.~\ref{Fig.1}: \begin{equation}
\label{ansatz}
u(x)= u_0 \exp{(-x/\ell)} \sin{(q x)}.
\end{equation} Substituting this into the our energy expression,  Eq.~(\ref{freeenergy}), we
determine $u_0$, $\ell$ and $q$ so as to minimize that energy. For applied loads less than a threshold given approximately by $f_c = 2 (\kappa \alpha_\perp)^{1/2}$, the minimum requires a straight MT: $u_0=q=0$. Above this threshold, the minimum is at a finite value of $u_0$ and $q\simeq1/\ell_0$. Once again, the buckling amplitude grows with applied load as $u_0 \sim (f- f_c)^{1/2}$ and decreases
with an increasing nonlinear elastic constant $\beta$ as $\beta^{-1/2}$ as found numerically (see Fig.~\ref{Fig.2}).
At a given load above threshold, $f > f_c$, the penetration depth of the oscillatory deformation field scales as $({\beta/\alpha_\parallel})^{1/2}$ (Fig.\ \ref{Fig.3}). The results of the variational calculation, therefore, corroborate all of the observations from the numerics.

We nondimensionlize the energy in Eq.~(\ref{freeenergy}) by
scaling $x$ and all displacements $u,v$ by $\ell_0$ --- e.g., $\tilde{u}(\tilde{x})= u(x/\ell_0)/\ell_0$. We also scale the energy by $f_0 \ell_0$, where $f_0=(\kappa \alpha_\perp)^{1/2}$.
The resulting dimensionless energy $\tilde{\mathcal{E}}=\frac{\mathcal{E}}{f_0 \ell_0}$ can be written as
\bea
\label{dimlessenergy}
\tilde{\mathcal{E}}= - \phi \tilde{v}(0) + \int _0^{\infty} \left[ {  \frac{1}{2}
{\tilde{u}^{''}(\tilde{x})}^2   +  \frac{1}{2} {\tilde{u}(\tilde{x})}^2 } \right ] d \tilde{x}  \nonumber \\
+  \int _0^{\infty} \left [ {\frac{\epsilon}{2} {\tilde{v}(\tilde{x})}^2  +  \frac{ \eta}{4} {\tilde{u}(\tilde{x})}^4} \right ] d \tilde{x},
 \eea
where $\phi = f/f_0$ is the dimensionless applied force and  $\tilde{x}=x/\ell_0$.  We have also introduced dimensionless material parameters $\epsilon= \alpha_{\parallel}/\alpha_{\perp}$ and $\eta= \beta {\ell_0}^2 /\alpha_{\perp}$. Based on the estimates above for $\alpha_{\parallel,\perp}$, we expect that $\epsilon<1/2$. The local strain is approximately given by $u'\simeq\tilde u$. For cytoskeletal networks, significant stiffening is expected for strains less than 1 \cite{Gardel,Storm}. Thus, we expect $\eta\gsim1$. Our calculations are all for $\eta/\epsilon \ge 10$, $0.5< \eta < 100$, and $0.001 < \epsilon <  0.1$.

\begin{figure}
\onefigure[width=8cm]{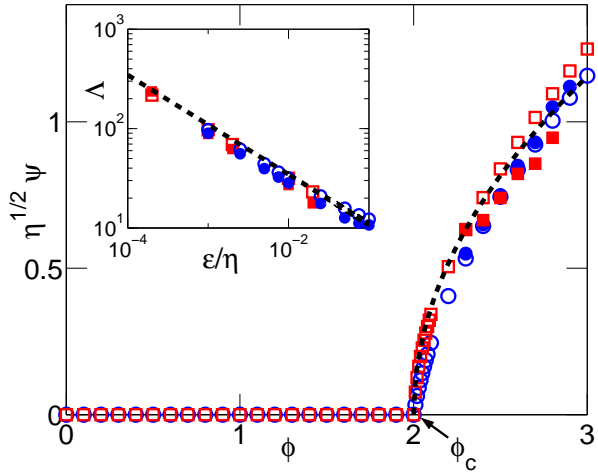}
\caption{Collapsed data from Figs 2 and 3. The main figure shows the scaled amplitude as a function of the dimensionless force $\phi$. The inset shows the dimensionless decay length $\Lambda$ as a function of $\epsilon/\eta$ the relative strength of the nonlinear terms for longitudinal and transverse coupling. Filled symbols correspond to the numerics, open symbols to the variational calculation, and the dashed lines to the asymptotic analysis.}
\label{Fig.4}
\end{figure}

In terms of the dimensionless variables, the ansatz in Eq.~(\ref{ansatz}) becomes $\tilde{u}(\tilde{x})= \psi \exp{(-\tilde{x}/\Lambda)} \sin{(\tilde{q} \tilde{x})}$, with $\Lambda=\ell/\ell_0$ and $\tilde{q} = \ell_0 q$. Using
this to minimize Eq.~(\ref{dimlessenergy}), we find the dependence of $\Lambda$ and $\psi$ on the applied load near threshold, $(\phi - \phi_c)/\phi_c\ll 1$, in the case of weak longitudinal coupling ($\epsilon \ll 1$), as follows. In this limit, $\Lambda\gg 1$ so we expand the energy Eq.~(\ref{dimlessenergy}) in $1/\Lambda$ and keep the leading terms, which are of order $\Lambda$ and $\epsilon\Lambda^3$. Dropping corrections that are suppressed by a factor of either $\epsilon$ or $1/\Lambda^2$, we determine the minimum of the truncated, asymptotic expansion of the energy with respect to $\psi$, $\Lambda$, and $\tilde{q}$. We find $\tilde{q}=q\ell_0=1$, and
\bea
\label{dimlessLambdapsi}
\Lambda = \sqrt{\frac{12 \eta}{\epsilon}}, \;\;\;\; \psi = \pm \sqrt{\frac{4}{3 \eta} (\phi-\phi_c)}.
\eea
To lowest order in $1/\Lambda$ and $\epsilon$, the critical force $f_c$ is unchanged from that of  the linear theory. Specifically, $\phi_c=2 + 4 \Lambda^{-2} + \cdots$. In Fig.~\ref{Fig.4} we show
results of the numerics (filled symbols), the variational analysis (open symbols) and the asymptotic analysis (lines) demonstrating the excellent agreement of these approaches.

\begin{figure}
\onefigure[width=8cm]{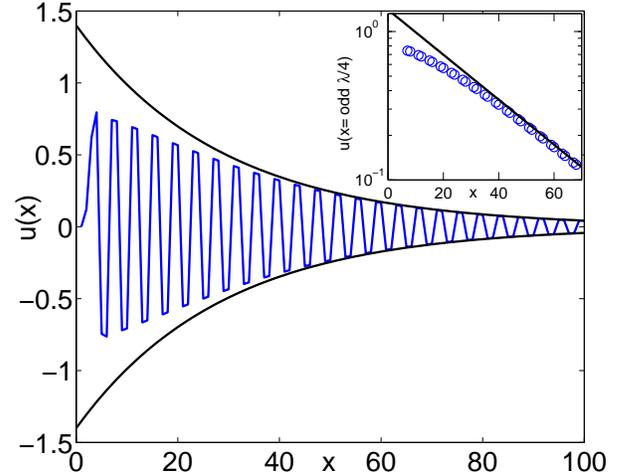}
\caption{The transverse displacement $u(x)$ in the buckled state of the MT in the presence of  higher order nonlinearity from numerics. The solid line (black) envelope corresponds to an asymptotic exponential ($1.4 \exp(-x/28.8)$) fit, with the same decay length as for the simple $u^4$ nonlinearity. The material parameters  are $\epsilon=0.01,\eta=1,\phi=3.5$. Inset: The circles (blue) show $u(x)$ at odd multiples of $\lambda/4$ (corresponding to the peaks in u(x)) as a function of axial length $x$ on a semilog scale for the higher order nonlinearity, while the solid line (black) shows the exponential decay with the simple $u^4$ nonlinearity.}
\label{Fig.5}
\end{figure}

The $\tilde{u}^4$ term in Eq.~(\ref{dimlessenergy}) represents only the initial nonlinear correction to the elasticity of the matrix. We expect a stronger nonlinear response for crosslinked cytoskeletal networks. Prior theory and experiment suggest an inverse quadratic divergence of the stress versus strain
 \cite{MacKintosh:95,Gardel,Storm} for F-actin networks. Motivated by this, we replace the transverse elastic terms in Eq.~(\ref{dimlessLambdapsi}) a term of the form $({\tilde u_s}^2/4) [| (\tilde{u}/\tilde u_s) + 1 |^{-1}   +  | (\tilde{u}/ \tilde u_s) -1 |^{-1}  - 2]$. This is chosen to reproduce the $\tilde u^2$ term in Eq.~(\ref{dimlessenergy}) to leading order. For small $\tilde u$, in fact, this should be equivalent to Eq.~(\ref{dimlessenergy}) with $\eta=2/\tilde u_s^2$. With this more physical nonlinearity, we find a clear departure from the exponential envelope for buckling oscillations observed earlier. In Fig.~\ref{Fig.5} we show the amplitude of the buckled MT for $\tilde u_s=1.4$; there appears to be a transition from a regime of slow amplitude decay near the point of loading to an exponential decay with the same rate as for the model in Eq.~(\ref{dimlessenergy}) with $\eta=1$.The inset compares the actual decay envelopes for the simple and higher order nonlinearities. The latter decreases the amplitude of buckling near the point of loading, however, the envelope asymptotes to exponential decay with a decay length equal to that for the simple nonlinearity of Eq.~(\ref{dimlessenergy}). The asymptotic approach to the exponential decay envelope (as predicted by
Eq.(\ref{dimlessenergy})) at smaller deformation amplitude is expected given that the elastic
energy shown in Eq.(\ref{dimlessenergy}) reproduces the lowest order terms of the more physical
elastic nonlinearities explored here.

Finally, we note that we have assumed the MT to be incompressible. Finite compressibility introduces another decay length;  the compressive strain on an elastic rod with compression modulus $\mu$ and longitudinal coupling to the
network of $\alpha_\parallel$ decays over a length scale $\sqrt{\mu/\alpha_\parallel}$ in the unbuckled limit. Treating a MT as an elastic rod of radius $a\simeq10$nm and Young's modulus $~1$GPa~\cite{dePablo}, we estimate this decay length to be  $\sim 10\mu$m for the largest  $\alpha_\parallel\simeq 1$kPa. At smaller $\epsilon$, where our our model is applicable to intracellular MTs, the actual decay length will be  larger than this, likely spanning the whole cell.

Motivated by observations of MT buckling in the cytoskeleton, we have
explored force propagation and compressive buckling of stiff rods
embedded in model nonlinear elastic matrices. Consistent with prior
experiments \cite{cliff}, we find oscillatory buckling with a decaying
amplitude and finite spatial extent $\ell$. We have focussed on the
limit of weak longitudinal coupling $\alpha_{\parallel}$, where the
penetration depth $\ell$ is large compared with the buckling wavelength
$\lambda$. We estimate these lengths using the experimental values from
Ref.~\cite{cliff}.  From $\kappa\simeq2\times10^{-23}$Nm$^2$ and
$G\simeq 1kPa$, we estimate that $\lambda\simeq 2\mu$m and
$\ell_0\simeq0.3\mu$m. The local strain in the network induced by a
dimensionless displacement $\tilde u$ is of order $\tilde u$.  Thus, for
a typical cytoskeletal network that becomes nonlinear for strains of
order 10-20\%, we estimate that $\eta$ is no less than unity, and may be
as large as 10. Thus, $\Lambda=\ell/\ell_0\gsim5$ since $\epsilon<1/2$.

Even for an isolated elastic rod, there are intrinsic nonlinearities
that arise from the geometry of a buckled shape \cite{reddick,winterflood}. For
small deflections, these nonlinearities can be obtained by expanding the
longitudinal field
$v(x)=\int_x^\infty\left(\sqrt{1+(u')^2}-1\right)$.
The leading correction to what we have considered above is a term of
order $(u')^4$, which appears in the first and next-to-last terms in Eq.(\ref{freeenergy}).
Within the limits that we consider here ($\epsilon<1, \Lambda>1$), the
dominant effect of these corrections is an additional term of order
$\tilde u^4$ in Eq.(\ref{dimlessenergy}). In detail, this geometric nonlinearity can be
captured within our asymptotic analysis by a simple substitution
$\eta \rightarrow \eta+\phi_c/2$  near the transition, which tends to decrease the
amplitude $\psi$ and increase the decay length $\Lambda$. But, it is
interesting to note that the existence of a decay length for the
buckling amplitude still depends crucially on the existence of finite
longitudinal coupling $\alpha_\parallel$.

We find that extrinsic nonlinearities of the embedding medium can result
in a continuous force--extension/compression relation, much as intrinsic
geometric nonlinearities can do for isolated elastic rods \cite{reddick,winterflood}. But, within
this model only longitudinal coupling of an elastic rod to a surrounding medium can
account for a decaying buckling amplitude, such as observed in the
experiments of Ref.\cite{cliff}. This work also shows how force transmission by
MTs can be controlled by the direct coupling of the MTs to the rest of
the cytoskeleton, which may be regulated by microtubule associating
proteins. This can allow an embedded MT to
bear and transmit supercritical forces over a long range within the
cell. Although we have considered here only the effects of the
confinement of MTs by and their coupling to a bulk elastic medium, it
is also possible that MTs near the plasma membrane may couple to the
cell substrate. Such coupling can also increase the load bearing
capacity of MTs. Force transmission such as we consider here is
different qualitatively from stress propagation in a homogenous and
isotropic elastic continuum in that the force is focussed primarily
along a single load bearing element. We find that this force decays
with a decay length $\ell$ to an asymptotic value $=f_c$. For typical
parameter values of $\eta=1$ and $\epsilon=0.01$, the lengthscale $\ell
\sim 35 \times \ell_0 \gsim 10 \mu\mbox{m}$.
Finite compressibility or elastic inhomogeneities in the surrounding
medium can also lead to finite--range force
transmission, although here we consider the case of  a single stiff
filament in an elastic continuum.
In this case we show that mechanical loads
applied at a cell's periphery can propagate along MTs throughout the cell
body without significant diminution. This elucidates the importance of
MTs in force propagation in a network of
F-actin with embedded MTs and agrees very well with recent experiments
\cite{YChia} that show that presence of MTs can significantly enhance the
elastic response of such a network.
While our study focusses on the mechanical response of intracellular
microtubules, our results are applicable to
composite elastic media with a wide separation of scales in stiffness of
its constituents such as biopolymer networks with bundles and even
nanotube-polymer composites.

\acknowledgments
MD  thanks C. Storm for helpful comments. FCM thanks C. Brangwynne for stimulating initial discussions. This work was supported in part by the Foundation for Fundamental Research on Matter (FOM) and NSF grant DMR-0354113.


\begin{thebibliography}{0}

\bibitem{alberts} 
\Name {Alberts B., Johnson A., Lewis J., Raff M., Roberts K. \and Walter P.}
\Book{Molecular Biology of the Cell}
\Publ{Garland Science, NewYork}
\Year{2005}.

\bibitem{cliff}
\Name{Brangwynne C.P., MacKintosh F.C., Kumar S., Geisse N.A., Talbot J.,
Mahadevan L., Parker K.K., Ingber D.E. \and Weitz D.A.}
\REVIEW {J. Cell Biology}{173}{2006}{733}.

\bibitem{ingber}
\Name{Alenghat F.J. \and Ingber D.E.}
\REVIEW{Science STKE}{119}{2002}{6}.

\bibitem{janmey:2004}
\Name{Janmey P.A. \and Weitz D.A.},
\REVIEW{Trends Biochem Sci.}{29}{2004}{364}.

\bibitem{janmey:1998}
\Name{P.A. Janmey}
\REVIEW{Physiological Reviews}{78}{1998}{763}.

\bibitem{wang}
\Name{Wang N., Butler J.P. \and Ingber D.E.}
\REVIEW{Science}{260}{1993}{1124}.

\bibitem{hu}
\Name{Hu S., Eberhardi L., Chen J., Love J.C., Butler J.P., Fredberg J.J., Whitesides G.M., and Wang N.}  
\REVIEW{Am. J. P. Cell Physiol.}{285}{2003}{C1082}.

\bibitem{landau}
\Name{Landau L.D. \and Lifshitz E.M.}
\Book{Theory of Elasticity}
\Publ{Pergamon Press, Oxford}
\Year{1986}.

\bibitem{skotheim}
\Name{Skotheim J.M. \and Mahadevan L.}
\REVIEW{Proc. R. Soc. Lond. A.}{460}{2004}{1995}.

\bibitem{dogterom} 
\Name{Dogterom M. \and Yurke B.}
\REVIEW{Science}{860}{1997}{278}.

\bibitem{press}
\Name{Press W.H., Teukolsky S., Vetterling W. \and Flannery B.}
\Book{Numerical Recipes in Fortran}
\Publ{Cambridge University Press}
\Year{1992}.

\bibitem{Crawford:91}
\Name{Crawford J.D.}
\REVIEW{Rev. Mod. Phys.}{63}{1991}{991}.

\bibitem{MacKintosh:95}
\Name{MacKintosh F.C., K\"{a}s J. \and Janmey P.A.}
\REVIEW{Phys. Rev. Lett.}{75}{1995}{4425}.

\bibitem{Gardel}
\Name{Gardel M.L., Shin J.H., MacKintosh F.C., Mahadevan L., Matsudaira P. \and Weitz D.}
\REVIEW{Science}{304}{2004}{1301}.

\bibitem{Storm}
\Name{Storm C., Pastore J., MacKintosh F.C., Lubensky T.C. \and and Janmey P.A.}
\REVIEW{Nature}{435}{2005}{191}.

\bibitem{dePablo} 
\Name{de Pablo P.J., Schaap I.A.T., MacKintosh F.C. \and Schmidt C.F.}
\REVIEW{Phys. Rev. Lett.}{91}{2003}{098101}.

\bibitem{reddick}
\Name{Reddick H.W. \and Miller  F.H.}
\Book{Advanced Mathematics for Engineers}
\Publ{John Wiley, New York}
\Year{1960}.

\bibitem{winterflood} 
\Name{Winterflood J., Blair D.G. \and Slagmolen}
\REVIEW{Phys. Lett. A}{300}{2002}{122}.

\bibitem{YChia} 
\Name{Lin Y.C., G. Koenderink, MacKintosh F.C. \and D. Weitz}
\REVIEW{unpublished}{ }{ }{ }.


\end{thebibliography}
\end{document}